# Global optimization for accurate determination of EBSD pattern centers

Edward L. Pang, Peter M. Larsen, Christopher A. Schuh[*]

*Department of Materials Science and Engineering, Massachusetts Institute of Technology, 77 Massachusetts Ave, Cambridge, MA 02139, USA*

*Corresponding author. Email address: schuh@mit.edu (C.A. Schuh)

**Highlights**

- Surprisingly large pattern shifts can be compensated by crystal rotations ("slop")
- Unavoidable noise is introduced by image processing and experimental conditions
- "Sloppy" and noisy landscape makes accurate pattern center determination difficult
- Use of global optimization and multiple patterns improves pattern center accuracy

**Abstract**

Accurate pattern center determination has long been a challenge for the electron backscatter diffraction (EBSD) community and is becoming critically accuracy-limiting for more recent advanced EBSD techniques. Here, we study the parameter landscape over which a pattern center must be fitted in quantitative detail and reveal that it is both "sloppy" and noisy, which limits the accuracy to which pattern centers can be determined. To locate the global optimum in this challenging landscape, we propose a combination of two approaches: the use of a global search algorithm and averaging the results from multiple patterns. We demonstrate the ability to accurately determine pattern centers of simulated patterns, inclusive of effects of binning and noise on the error of the fitted pattern center. We also demonstrate the ability of this method to accurately detect changes in pattern center in an experimental dataset with noisy and highly binned patterns. Source code for our pattern center fitting algorithm is available online.

# 1. Introduction

Electron backscatter diffraction (EBSD) relies on obtaining backscattered electron diffraction patterns, also known as Kikuchi patterns, and the manner of analyzing and indexing those patterns lies at the heart of the most pressing technological limitations of the technique, including accuracy and speed. The most traditional approach is to index patterns with the aid of a mathematical technique known as the Hough or Radon transform. The advantage of this approach is its speed—patterns can be indexed on-the-fly by the acquisition software. With this gain in speed, however, comes a trade-off in angular resolution—Hough-based indexing is commonly cited as having a precision of ~0.5° [1–3] and an absolute accuracy of 1-2° [4,5]. Nonetheless, this level of accuracy is adequate for many applications, and this is the method currently implemented by all major EBSD packages. More recently, advanced indexing techniques such as pattern matching [6] and dictionary indexing [7,8] have been proposed as alternatives to Hough-based indexing. These methods sidestep the Hough transform (and its associated error) and instead directly compare intensity information from the entire pattern against dynamically simulated patterns from forward modeling [9,10]. This type of approach has been shown to give more precise orientation measurements [11,12] and is also highly robust against noisy patterns [8,13].

There is a longstanding issue that has always contributed to limiting the accuracy of EBSD and which is becoming ever more relevant as the indexing methods improve and eliminate other errors. That issue is the ability to accurately determine the pattern center [14–16], which describes the position ($X^*$, $Y^*$, $Z^*$) of the sample relative to the EBSD detector and is required for pattern simulations. Typically, the pattern center is established using an iterative fitting algorithm first proposed by Krieger-Lassen *et al.* [17,18], which involves identifying band positions via the Hough transform and then altering the pattern center to minimize the difference between the measured and expected interplanar angles. This method has been shown to give a sensitivity of ~0.5% of the detector width [17–19], which is sufficient for standard EBSD but not for more advanced applications such as measuring strain (high-resolution EBSD, HR-EBSD) [20,21] or resolving pseudosymmetry [15,22,23]. As improvements on this approach, a number of clever in-microscope techniques based on moving screens [24–26], shadow casting [27–29], and known calibrants [19,30,31] have been developed over the years, but these methods suffer from drawbacks such as increased data acquisition time, inconsistent camera insertion, special hardware and modifications to the microscope, and inability to reposition sample height accurately owing to the significant depth of field in the SEM. It would therefore be preferable to use an entirely offline method following Krieger-Lassen *et al.*

More recently, a number of improved offline techniques have been suggested that employ simulated patterns via forward modeling, such as a strain minimization technique described by Fullwood *et al.* that is implemented in the open-source HR-EBSD package OpenXY [14] and the optimization routines in the dictionary indexing software EMsoft [8], among others [32,33]. These new methods, like the dictionary indexing and pattern matching techniques, take advantage of the enhanced sensitivity offered by the use of forward modeling and have led to significant improvements in pattern center accuracy, possibly within ~0.03-0.05% of the detector width that is claimed to be necessary for absolute strain values to be obtained in HR-EBSD [15,26,34,35]. For example, one promising development is the observation that averaging the results from several patterns can reduce some types of error [14,15,17,32,36]. However, there remains a general problem that these algorithms attempt to simultaneously extract pattern center, orientation, and possibly strain information from a single pattern, and it is unclear to what extent these quantities can accommodate or mask each other. With this in mind, Tanaka and Wilkinson recently proposed the use of a differential evolution global optimization algorithm to simultaneously fit pattern centers and orientations [33]. However, they did not investigate the optimization landscape, the nature of which will determine the accuracy limits of this technique, or evaluate the potential accuracy of their method on real experimental patterns.

In our recent work on resolving pseudosymmetry in zirconia [23], we also pointed to the need for more accurate pattern center estimates and a better understanding of the potential errors from a forward modeling based fitting approach. It is the purpose of the present paper to present our method in greater detail and quantitatively study the problem of simultaneously optimizing the pattern center and crystal orientation by comparison to dynamically simulated patterns. Our results reveal aspects of the optimization landscape that make accurate determination of the pattern center difficult, which justify the use of two key components combined for the first time on this problem: a global search algorithm and the use of multiple patterns. We then test our method on simulated and experimental datasets, comparing its accuracy and precision to existing algorithms in the literature.

## 2. Optimization landscape

### 2.1. Defining the optimization problem

We follow the approach of forward modeling through physics-based simulation of EBSD patterns given a set of parameters that describe the sample and detector [9,10]. The objective of this approach is to find parameters for the forward model such that the resulting simulated pattern has the best possible match with the experimental pattern. The parameters that we aim to fit are the pattern center (three parameters) and the orientation of the crystal (three parameters). We assume here that the crystal structure is known *a priori*. There are numerous other parameters such as camera tilt angle and gamma correction factor that must also be specified, but these are usually considered fixed and are not optimized. As such, we will not consider them further. Interested readers are referred to refs. [8,37] for more details.

The fitting process usually proceeds in the following manner: 1) an initial pattern center and orientation are specified and used to simulate a test pattern, 2) this test pattern is compared to an experimental pattern via some similarity metric, 3) the pattern center and orientation are both altered in an attempt to improve the similarity metric, and 4) this process proceeds iteratively until a satisfactory pattern center and orientation are obtained. We note that the orientation must simultaneously be optimized with the pattern center, as it is known that incorrect orientations can lead to spurious shifts in the best-matching pattern center [36,38]. A schematic of this process in a simplified two-dimensional space is shown in Fig. 1, which demonstrates how the simulated test pattern evolves as the search algorithm moves towards the optimal orientation and pattern center.

In this work, we use the normalized dot product (NDP) as the similarity metric between patterns in the spirit of the dictionary indexing approach pioneered by Chen *et al.* [7] and implemented in the open-source EMsoft package [8]. The NDP between the experimental pattern $\boldsymbol{A}$ and test pattern $\boldsymbol{B}$ is computed as follows:

$$\text{NDP} = \frac{\sum_{i,j} A_{ij} B_{ij}}{\|\boldsymbol{A}\| \|\boldsymbol{B}\|} \tag{1}$$

where $\|A\| = \sqrt{\sum_{i,j} A_{ij}^2}$ and $A_{ij}$ is the intensity of image $A$ at pixel location $(i,j)$. The NDP can vary from 0 to 1, with higher values representing a greater degree of similarity between the images.

Three parameters are needed to fully specify orientation, and here, we represent misorientations in terms of a three component Rodrigues vector $R$ given by:

$$R = [R_x, R_y, R_z] = \tan\left(\frac{\alpha}{2}\right)\hat{r} \qquad (2)$$

where $\hat{r}$ is a unit vector describing the axis of rotation and $\alpha$ is the angle of rotation about this axis [39]. For small misorientation angles, given by $R$ near $[0, 0, 0]$, this space contains no singularities and is essentially uniform (unlike Euler angles [18]) and has no constraints on its parameters (unlike unit quaternions). These properties allow us to apply standard unconstrained optimization methods to maximize the NDP in a six-dimensional space of $X^*$, $Y^*$, $Z^*$, $R_x$, $R_y$, and $R_z$.

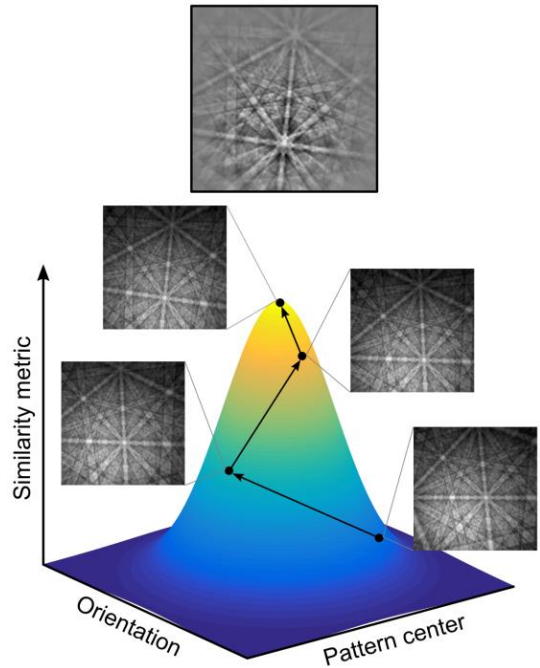

**Fig. 1.** Schematic of the present optimization problem, showing a hypothetical search path towards the peak similarity metric by varying orientation and pattern center. The experimental pattern is shown at the top whereas the bottom four patterns were simulated by forward modeling. A simplified two-dimensional space is shown here for clarity, but the full optimization occurs in six-dimensional space encompassing three pattern center parameters and three orientation parameters.

## 2.2. "Slop" and noise in the optimization landscape

It has been previously pointed out that small shifts in the pattern center can be accommodated by, and are almost indistinguishable from, small crystal rotations [36]. To quantify this effect, we have evaluated the similarity of patterns subjected to changes in $Y^*$ and $y$-rotation, which represent shifting the crystal (or detector) vertically and rotating the crystal in a direction that moves the EBSD pattern in the vertical direction, respectively (Fig. 2a). A simulated pattern of α-Fe was taken as the "experimental" pattern, and the NDP landscape was mapped out in this two-dimensional parameter space (similar to that shown in Fig. 1) by comparing with simulated patterns of varying $Y^*$ and $y$-rotation, holding all else constant.

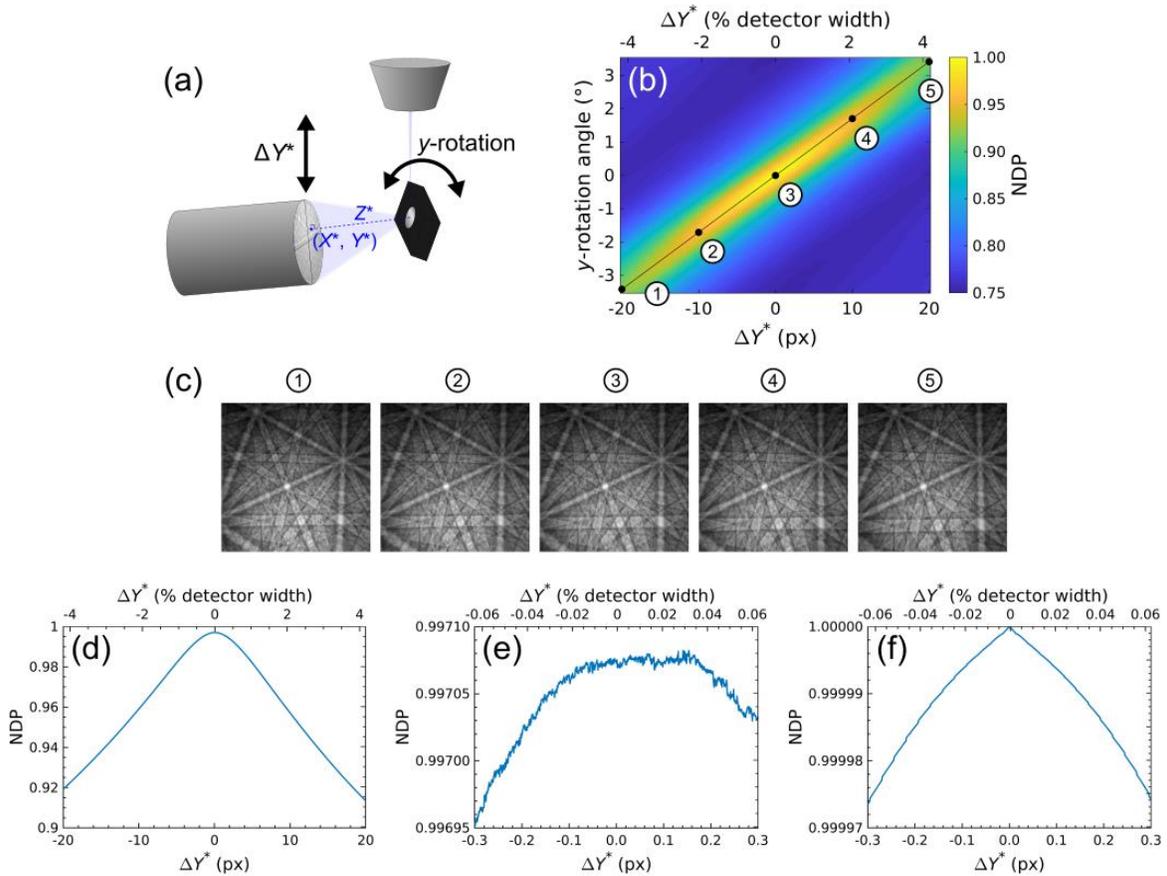

**Fig. 2.** Demonstration of slop and noise in the optimization landscape. (a) Schematic illustrating the crystal rotation and pattern shift directions for the data shown in this figure. (b) NDP as a function of $\Delta Y^*$ and $y$-rotation from the true pattern center and orientation on simulated patterns of α-Fe. (c) Simulated patterns corresponding to the points shown in part b, illustrating that similar patterns can be obtained even with large pattern shifts after a compensating crystal rotation. (d-e) Linescans along the ridge in part b at low (d) and high (e) magnifications with image processing employed in the EMsoft suite. (f) Same linescan as in part e but without any image processing.

Patterns of α-Fe with size 480×480 px (pixel size 50 μm) were simulated using the EMEBSD program within EMsoft 4.0 [8]. Simulation parameters are given in Table 1. NDP values were computed using EMEBSDDI, the dictionary indexing program within EMsoft, employing a circular mask with diameter equal to the image width, a high-pass filter parameter of 0.05, and adaptive histogram equalization with 10 regions. We chose the following pattern center to represent a typical microscope geometry:

$X^* = 0$ px (0.500000 in fraction detector width)

$Y^* = 80$ px (0.666667)

$Z^* = 15000$ μm (0.625000).

In this work, when ($X^*$, $Y^*$, $Z^*$) are given in fraction or percent detector width, these are in reference to the EDAX/TSL coordinate system. We also give these quantities in units of px for $X^*$ and $Y^*$ and μm for $Z^*$, which refer to the EMsoft convention (sometimes called $x_{pc}$, $y_{pc}$, and $L$, respectively). For further explanation of these different conventions, see Ref. [8].

**Table 1.** *EMsoft* parameters used for the simulated α-Fe dataset in the present study.

| *Monte Carlo and master pattern simulation parameters* | | *Pattern simulation parameters* | |
|---|---|---|---|
| Total number of incident electrons | 2×10⁹ | Camera elevation | 5˚ |
| Specimen tilt angle | 70˚ | Incident beam current | 150 nA |
| Incident beam energy | 20 keV | Beam dwell time | 100 μs |
| Minimum BSE exit energy | 5 keV | Gamma value | 0.33 |
| Energy bin size | 1 keV | Detector size | 480×480 px |
| Maximum exit depth | 100 nm | Detector pixel size | 50 μm |
| Depth step size | 1 nm | Bit depth | 8 bit |
| Smallest d-spacing | 0.05 nm | | |
| Master pattern size | 1001×1001 px | | |
| Material | α-Fe (bcc) | | |
| Lattice parameter | 0.28665 nm [40] | | |
| Debye-Waller factor | 0.003106 nm² [41] | | |

The NDP as a function of $Y^*$ and *y*-rotation is shown in Fig. 2b, which reveals a highly elongated peak. The long axis of this peak reveals that changes in $Y^*$ of 6 px (1.25% detector width) can be compensated by rotations of ~1°, leading to high NDP values over a wide range of $Y^*$. Physically, this means that highly similar EBSD patterns can be generated from a wide range of $Y^*$ and *y*-rotation values, as seen in Fig. 2c.

This phenomenon is known as "slop" and has been well-studied in other fields such as systems biology [42,43]. Sloppy models are ones that depend on variables that can compensate for each other, leading to similar model outcomes for a wide range of input parameters. From a geometric perspective, this corresponds to an optimization landscape that has one or more directions with small gradients near the global optimum. As a result, collectively fitting parameters by optimizing the agreement between the model and experimental data can lead to large parameter uncertainties.

In a sloppy landscape, the effect of noise can be extremely detrimental to the already large parameter uncertainties. In addition to riddling the landscape with local optima in which the optimization algorithm can get trapped, noise can have an additional effect of displacing the true optimum of the landscape. To illustrate, Fig. 2d-e shows the variation in NDP along the ridge in Fig. 2b. While the linescan appears smooth when looking at large changes in NDP (Fig. 2d), upon closer examination it is apparent that this peak is rather noisy for the small changes in pattern center on the order of 0.01% in which we are interested (Fig. 2e). As can be seen, the point of maximum NDP is not coincident with the true pattern center, and the $Y^*$ giving the highest NDP value is actually 0.15 px (0.03% detector width) away from the true $Y^*$. This can be explained by the flatness of the landscape near the peak, owing to slop, which makes it easier for noise to elevate the NDP of surrounding regions above that of the true optimum.

The noise in this landscape originates from the digital image processing steps taken to normalize the image intensities, namely high-pass filtering and adaptive histogram equalization, as the same linescan taken without any image processing (Fig. 2f) is essentially smooth; it contains only very small steps at regular intervals arising from the discrete nature of digital images, which would not give standard optimization algorithms much difficulty. However, that standard image processing steps introduce noise to the landscape to the degree seen in Fig. 2e is, we believe, an underappreciated point in the EBSD community that warrants future investigation. Critically, such image processing steps are absolutely required for experimental patterns to ensure that the NDP does not pick up spurious effects. When testing any optimization method for pattern center matching, such steps should therefore be included, although many papers on this topic either do not include, or are unclear on whether they include, such essential image processing steps. In everything that follows in this paper, we include these detrimental image processing steps to ensure that our tests on simulated patterns do not give misleadingly accurate results.

## 3. Performance of the Nelder-Mead simplex algorithm

The above results illustrate the fundamental difficulty faced in the present optimization problem; even if a perfect optimization algorithm were available, correct parameters cannot be found due to the combined effect of slop and noise in displacing the optimum from the true pattern center. These findings suggest a practical limit to the accuracy that the pattern center can be determined from a single EBSD pattern by an iterative fitting procedure that performs a comparison with dynamically simulated patterns. While we have demonstrated these issues on a reduced two-dimensional parameter space for simplicity and ease of visualization, these same issues persist and are exacerbated in the full six-dimensional space, as we will develop in more detail below. In this section, we explore the use of a local optimization algorithm and the limits of its abilities to find the pattern center in light of the sloppy and noisy optimization space.

*3.1. Fitting on a single pattern*

We first evaluate the performance of the popular Nelder-Mead simplex algorithm [44], which has been used for EBSD data refinement by other authors [6,8], for optimizing the pattern center and orientation on a single EBSD pattern. We used the standard MATLAB implementation of the Nelder-Mead algorithm, *fminsearch*, with a stopping tolerance of $1\times10^{-6}$ on the NDP. All variables were centered and scaled by factors of 9000 px for $X^*$ and $Y^*$, 900000 μm for $Z^*$, and 15 for Rodrigues vectors, which were found to give optimal performance. These scaling factors led to an initial simplex step size of 2.25 px in $X^*$ and $Y^*$, 225 μm in $Z^*$, and 0.43° in $R_x$, $R_y$, and $R_z$ (in non-scaled units).

Patterns of α-Fe (480×480 px) were simulated for 10 random orientations using the same methods given in Section 2.2. The starting pattern center values for optimization were perturbed from the true values:

$$X^* = -2.3664 \text{ px } (-0.49\% \text{ detector width})$$

$$Y^* = 78.8267 \text{ px } (-0.24\%)$$

$$Z^* = 15135.51 \text{ μm } (+0.56\%).$$

The starting orientation for optimization was determined by perturbing each Euler angle by a random amount within ±1° from the true orientation.

After optimization, surprisingly large errors from the true pattern center were obtained, up to 1.3 px (0.26% detector width) in $X^*$, 1.4 px (0.29%) in $Y^*$, and 23 μm (0.10%) in $Z^*$. Interestingly, the errors from the Nelder-Mead algorithm are not centered about zero. In fact, they are biased towards the starting values, which suggests that the Nelder-Mead algorithm terminates prematurely in a local optimum, as commonly reported [45–47]. Confirming this point, the NDP for the fitted solution is on average $3.2 \times 10^{-4}$ (and at most $1.1 \times 10^{-3}$) lower than for the actual solution, which considering the landscape shown in Fig. 2e demonstrates that the fitted solution is quite far from the optimum. This demonstrates that using a single EBSD pattern to fit the pattern center is not sufficient to obtain high levels of accuracy.

*3.2. Averaging over multiple patterns*

In an attempt to improve the accuracy of the pattern center estimate, we investigated the effect of averaging over multiple patterns, as has been suggested in other EBSD papers using local optimization routines [14,15,17,32,36]. Fig. 3 shows the average pattern center for all possible subsets of the 10 patterns to demonstrate the effect of averaging on the resulting pattern center estimate. Each of the $\binom{10}{N}$ points represents the average pattern center considering a single combination of $N$ patterns. Thus, the 10 points for $N = 1$ represent the pattern centers determined from the 10 individual patterns, and the single point for $N = 10$ represents the average over all 10 patterns. As $N$ increases, the average pattern center clearly improves; after averaging over 10 patterns, the worst error ($X^*$) was reduced to 0.07% detector width. These results demonstrate the need to average over multiple patterns for the best possible accuracy.

However, even so, the averaged pattern center does not converge towards the true value as more patterns are included because of the biased estimates found by the Nelder-Mead algorithm. Thus, while averaging over numerous grains improves the pattern center estimate, its accuracy is more fundamentally limited by the inability of local search algorithms such as Nelder-Mead simplex to cope with the challenging optimization landscape. To further improve the accuracy of pattern center determination, we need another search algorithm that is better suited for the present problem.

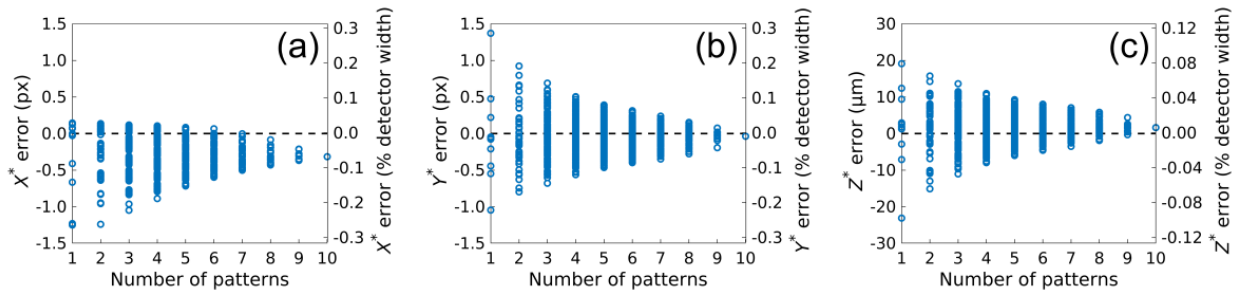

**Fig. 3.** Pattern center errors in (a) $X^*$, (b) $Y^*$, and (c) $Z^*$ as a function of the number of patterns averaged for 10 simulated patterns of α-Fe (480×480 px) as determined by the Nelder-Mead simplex algorithm.

## 4. Use of a global search algorithm for improved pattern center determination

Having established that local search algorithms such as Nelder-Mead simplex cannot produce highly accurate pattern center estimates, even after averaging over multiple EBSD patterns, we instead propose the use of a global search algorithm to optimize the objective function. Such an algorithm, when combined with the use of multiple EBSD patterns, will be shown to produce the best possible accuracy.

### 4.1. Performance of the SNOBFIT algorithm

We have selected the SNOBFIT (Stable Noisy Optimization by Branch and Fit) algorithm [48], which belongs to a class of derivative-free global optimization algorithms [46]. To avoid getting trapped in local optima, it not only moves in the direction of improving objective but also evaluates unexplored space with some probability during each iteration. We used a MATLAB implementation of the SNOBFIT algorithm, which was downloaded from Ref. [49]. All variables were scaled and centered in the same manner as for Nelder-Mead. The SNOBFIT parameters used are given in Table 2.

**Table 2.** SNOBFIT parameters used in the present study.

| Parameter | Value |
|---|---|
| Number of random start points, $n_{point}$ | 12 |
| Number of points generated in each iteration, $n_{req}$ | 12 |
| Probability of a random search point, $p$ | 0.5 |
| Minimum step size in $X^*$, $Y^*$ | 0.0025 px |
| Minimum step size in $Z^*$ | 0.25 µm |
| Minimum step size in $R_x$, $R_y$, $R_z$ | 0.0005° |
| Trust radius for $X^*$, $Y^*$ | 5 px |
| Trust radius for $Z^*$ | 500 µm |
| Trust radius for $R_x$, $R_y$, $R_z$ | 1° |
| Maximum function evaluations | 1000 |

For a direct performance comparison, we have conducted the same tests as for the Nelder-Mead algorithm using the same 10 patterns and optimization starting points, and the corresponding SNOBFIT results are shown in Fig. 4. The data clearly show a significantly reduced spread—the pattern center errors on a single EBSD pattern are on the order of 0.05% detector width, significantly lower than obtained for Nelder-Mead (~0.2%). In addition, the NDP for the fitted solution is on average only $3.1 \times 10^{-5}$ lower than for the actual solution, an order of magnitude better than for Nelder-Mead and approaching the noise level in the landscape (~$10^{-5}$ as seen in Fig. 2e). In addition, half of the points fitted by the SNOBFIT algorithm found a higher NDP for the fitted solution compared to the ground truth. Because of the stochastic nature of SNOBFIT, the algorithm would likely find a higher NDP for all points if the number of allowed function evaluations was increased, which would further improve the accuracy at the expense of additional computation time. This gives confidence that the SNOBFIT algorithm generally finds the global optimum and confirms that the global maximum is not necessarily coincident with the true pattern center, which fundamentally limits the possible accuracy to which the pattern center can be fit for a single EBSD pattern (~0.05% of the detector width for the present microscope setup).

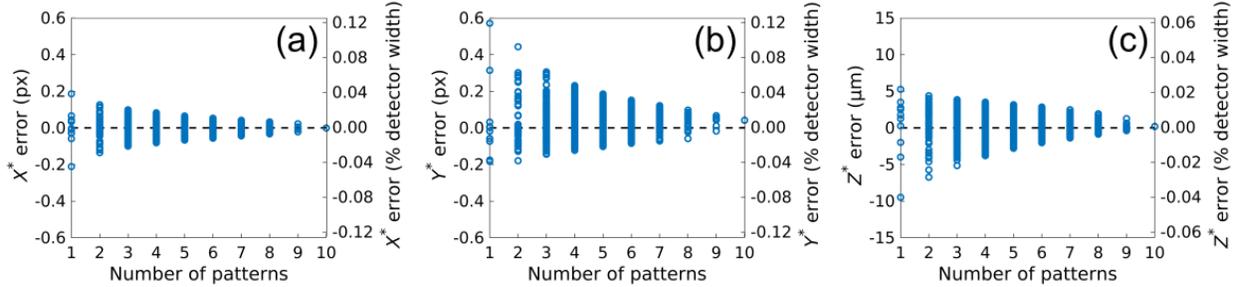

**Fig. 4.** Pattern center errors in (a) $X^*$, (b) $Y^*$, and (c) $Z^*$ as a function of the number of patterns averaged for 10 simulated patterns of α-Fe (480×480 px) as determined by the SNOBFIT algorithm. Note the different scales compared to Fig. 3; the present y-scales are magnified to accommodate the substantially smaller errors obtained by the SNOBFIT algorithm.

After taking the average pattern center over 10 patterns, highly accurate pattern centers are obtained, with errors in the range of 0.0002-0.008% detector width, about an order of magnitude more accurate than that found by Nelder-Mead simplex (a side-by-side comparison is given in Table 3). Thus, not only does moving to a global search algorithm significantly improve the pattern center estimate for a single pattern, it reduces the bias of the estimates, which additionally enhances the effectiveness of averaging over multiple patterns in extracting the true pattern center. These results demonstrate the effectiveness of combining global optimization with the use of multiple EBSD patterns for the best possible accuracy of pattern center determination.

**Table 3.** Mean and standard deviation of the average pattern center errors and ΔNDP (NDP of the fitted solution minus that of the actual solution) using the Nelder-Mead simplex and SNOBFIT algorithms for 10 simulated patterns of α-Fe.

|  | $X^*$ | $Y^*$ | $Z^*$ | ΔNDP |
| --- | --- | --- | --- | --- |
| Nelder-Mead simplex | -0.32 ± 0.55 px (-0.066 ± 0.115%) | -0.04 ± 0.65 px (-0.008 ± 0.135%) | 1.6 ± 11.5 μm (0.007 ± 0.048%) | -3.2 ± 4.3×10$^{-4}$ |
| SNOBFIT | -0.0009 ± 0.10 px (-0.0002 ± 0.021%) | 0.04 ± 0.23 px (0.008 ± 0.048%) | 0.2 ± 4.3 μm (0.0008 ± 0.018%) | -3.1 ± 6.9×10$^{-5}$ |

*4.2. Effect of binning and noise*

Thus far, we have only considered high-quality simulated patterns without binning and noise. However, experimental patterns are noisy and typically binned to smaller image sizes, which affects the accuracy of the fitted pattern center. To study the effect of binning and noise, we added varying amounts of noise to the same 10 simulated patterns of α-Fe from Section 3 and subjected them to different levels of binning. Gaussian noise was added using the *imnoise* function in MATLAB. Variances of 0.03, 0.15, and 0.75 (in fraction of the intensity range) were used to generate images with peak signal-to-noise ratios of approximately 15.6, 10.3, and 7.4 dB, respectively. The patterns were then binned in MATLAB by a factor of 1, 2, 4, or 8. Representative images are shown in Fig. 5. For each combination of binning and noise, we performed the same tests as in Fig. 4 using the SNOBFIT algorithm. The resulting mean pattern center errors over the 10 patterns are summarized in Fig. 6a-c, and the standard deviations of the errors are given in Fig. 6d-f.

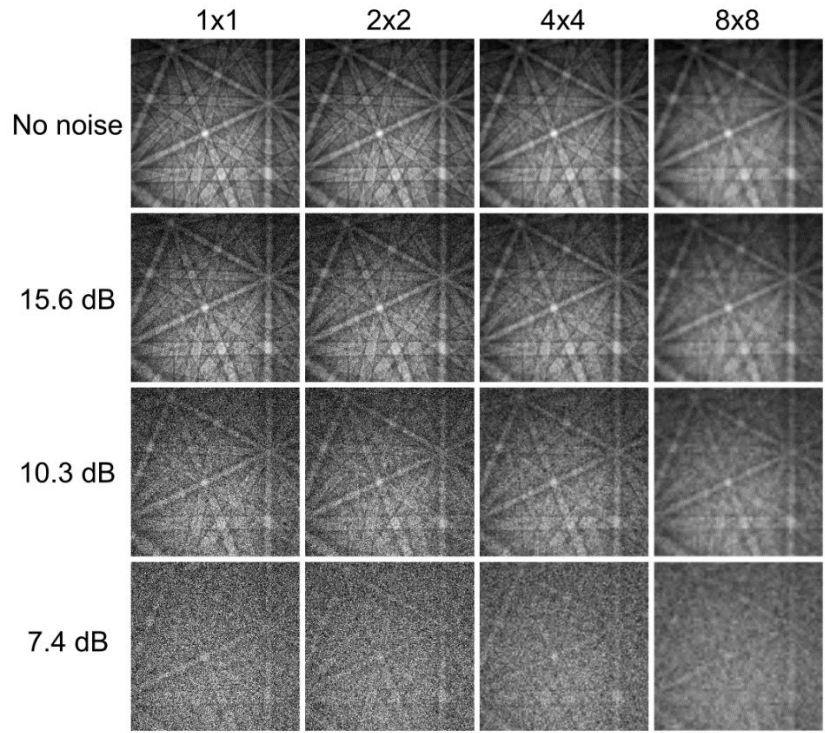

**Fig. 5.** Simulated patterns of α-Fe with various levels of noise and binning used in this investigation. Noise level is quantified by the peak signal-to-noise ratio for the unbinned (1×1) pattern. Unbinned image size is 480×480 px.

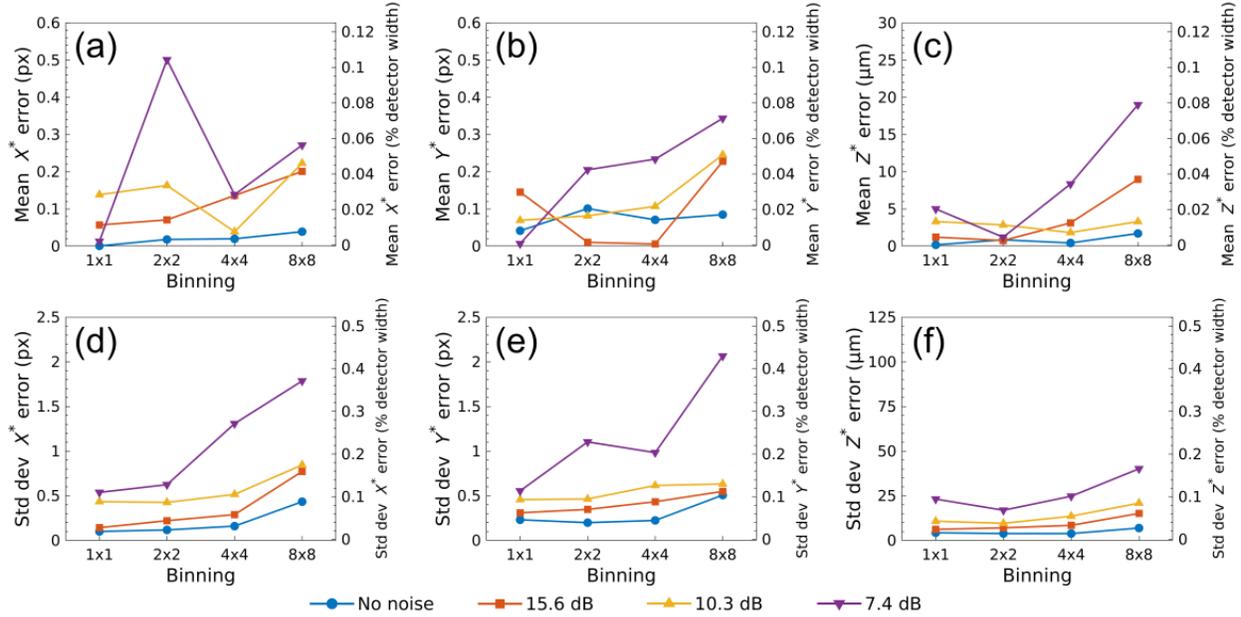

**Fig. 6.** Accuracy and precision of the pattern center determined from 10 simulated patterns of α-Fe using the SNOBFIT algorithm as a function of binning and noise. (a-c) Absolute value of the average pattern center error. (d-f) Standard deviation of the pattern center error.

These results show that accuracy and precision both generally degrade with an increasing level of noise. However, the accuracy of the SNOBFIT pattern center is surprisingly tolerant to binning and noise. For example, the patterns without additional noise added show no clear increase in the mean or standard deviation going from 1×1 to 4×4 binning and only a small increase going to 8×8 binning. Even so, the average pattern center for noise-free patterns with 8×8 binning is fitted to within 0.02% detector width from the true value, which is rather remarkable considering that this image has dimensions of only 60×60 px. A similar trend emerges as noise is added to the unbinned patterns, and the noisiest patterns with a peak signal-to-noise ratio of 7.4 dB have mean errors within ~0.03% detector width. With high levels of both noise and binning, there is a more marked deterioration in the accuracy of the determined pattern center. Even so, it appears that the pattern center can be determined to within ~0.1% detector width for the most noisy and highly binned patterns after averaging the pattern center from 10 patterns.

The reason for the increased error for noisy and binned patterns is the same clouding of the true maximum previously demonstrated in Fig. 2e and undoubtedly increases the likelihood of early stopping in a local optimum for standard search algorithms; this is why previous studies have focused on high-quality unbinned patterns for pattern center determination [8,15,32,34]. However, the fact that SNOBFIT can determine the pattern center to subpixel accuracy even for noisy and highly binned patterns demonstrates the ability of this algorithm to successfully navigate the challenging landscape and extract the pattern center to good levels of accuracy. It also demonstrates the robustness to noise and binning of whole pattern intensity comparisons to dynamically simulated patterns, as previously documented for forward-modeling based EBSD indexing techniques [8,13].

*4.3. Comparison to existing algorithms*

The present SNOBFIT results compare favorably to existing pattern center determination algorithms. Here, we only consider similar approaches that involve iterative simulations of dynamical diffraction patterns to achieve the best match against a reference pattern. Other promising approaches have been reported in refs. [32,34,50].

Jackson *et al.* [15] used a strain minimization approach [14] that involves comparing the test pattern with dynamically simulated patterns with varying pattern centers to determine the best match, as in the present study. On a simulated dataset of 64 high-quality Si patterns of size 1024×1024 px, they found errors of 0.007 ± 0.041% detector width for $X^*$, 0.008 ± 0.024% for $Y^*$, and 0.016 ± 0.019% for $Z^*$. However, the present results using only 10 patterns of size 480×480 px obtained even lower average errors of 0.0002-0.008%. While a direct comparison is difficult owing to differences in material, orientations, and starting points for the optimization, the fact that we obtained a better average accuracy with significantly fewer and smaller patterns suggests that the SNOBFIT algorithm more effectively finds the true pattern center. This is likely a consequence of the present method only optimizing in the six-dimensional space of pattern center and orientation, whereas the strain minimization technique considers the extra degrees of freedom presented by strain during cross-correlation, even when no strain is present in the test or reference patterns. As strain can be confounded with rotations and pattern shifts [15,51], this exacerbates the problem of slop and reduces the possible accuracy that can be attained. Based on our results and those of Jackson *et al.*, we are optimistic about the use of global optimization algorithms for improving results on the larger problem including strain extraction.

Tanaka and Wilkinson have taken a similar approach as the present study that involves direct optimization of pattern center and orientation in a six-dimensional space [33]. They used a differential evolution global search algorithm combined with dynamical patterns simulated by Bruker DynamicS software and used the cross-correlation coefficient as their similarity metric. For simulated patterns of α-Fe of size 478×478 px, comparable to our 480×480 px patterns, they obtained an average error from 10 patterns of approximately 0.001% for $X^*$, 0.003% for $Y^*$, and 0.001% for $Z^*$, similar to the present errors of 0.0002-0.008%. While they also found that the addition of noise to their simulated patterns reduced the accuracy of the fitted pattern centers, they found that binning did not, which is contrary the present results and not immediately intuitive. One possible explanation is different image processing procedures; as we noted earlier, our use of the standard high-pass filtering and adaptive histogram equalization introduces noise to the optimization landscape. It is possible that the work in Ref. [33] may have used different procedures leading to a smoother landscape (cf. Fig. 2f) than the present study (cf. Fig. 2e). If our optimization landscape were noisier, it could explain why our results are more sensitive to binning than those of Tanaka and Wilkinson. However, it is important to note again that our image processing steps are intentionally chosen to produce such noise, because such steps are unambiguously needed for experimental work. We therefore turn to experimental data for our final analysis in the next section.

## 5. Case study on an experimental Ni dataset

In order to demonstrate the validity of our approach under real conditions, we evaluate its performance on the Ni-1 dataset published alongside Ref. [8]. The inverse pole figure (IPF) map for Hough-indexed data is shown in Fig. 7a. This dataset contains highly binned patterns of size 60×60 px, an example of which is shown in Fig. 7b. These images are both noisy and aggressively binned but represent typical fast mapping conditions. Thus, this dataset provides a useful test case to evaluate the effectiveness of the SNOBFIT algorithm.

For pattern center fitting using SNOBFIT, we used the master pattern provided with the dataset. A gamma value of 0.33 was used with a high-pass filter parameter of 0.05 and adaptive histogram equalization with 10 regions. A circular mask of radius 30 px was used for all NDP calculations. The same SNOBFIT parameters given in Table 2 were used. For optimization, the pattern center from the Hough-indexed data was used as the starting pattern center for all map points:

$$X^* = 3.4858 \text{ px } (0.507262)$$

$$Y^* = 114.2035 \text{ px } (0.737924)$$

$$Z^* = 15870.02 \text{ μm } (0.558489).$$

The starting orientations were generally the Hough-indexed orientations. For the few misindexed points, the orientation of a correctly indexed neighboring point was used instead.

To test the ability of the SNOBFIT algorithm to detect small changes in pattern center, we have fitted the pattern center along the vertical line in Fig. 7a. Because of the large height of the region (225 μm), a significant pattern center shift is expected as the beam scans across this line. Expected changes in $Y^*$ and $Z^*$ with changes in map $y$-position are given by:

$$\frac{\Delta Y^*}{\Delta y} = \cos(90° - \sigma + \theta_c)$$

$$\frac{\Delta Z^*}{\Delta y} = \sin(90° - \sigma + \theta_c) \qquad (3)$$

where $\sigma$ is the sample tilt angle and $\theta_c$ is the camera elevation angle, which are 75.7° and 10°, respectively, for this dataset. The goal of this test is to see if SNOBFIT can accurately detect these changes.

For each $y$-position in the map, we have fitted the pattern center to the 10 patterns in the width of the shaded band in Fig. 7a, and the results of this test are given in Fig. 7c-e. In this linescan, we are moving in the positive $y$-direction, so $Y^*$ and $Z^*$ should increase whereas $X^*$ should remain constant. The results shown in Fig. 7c-e are in good agreement with expectation. In addition, the solid lines drawn in the plots, which correspond to the theoretically expected change in pattern center calculated from Eq. (3), show excellent agreement with the slope of the data points. Thus, these results demonstrate that the SNOBFIT algorithm can accurately detect relative changes in pattern center, even for 8×8 binned experimental patterns.

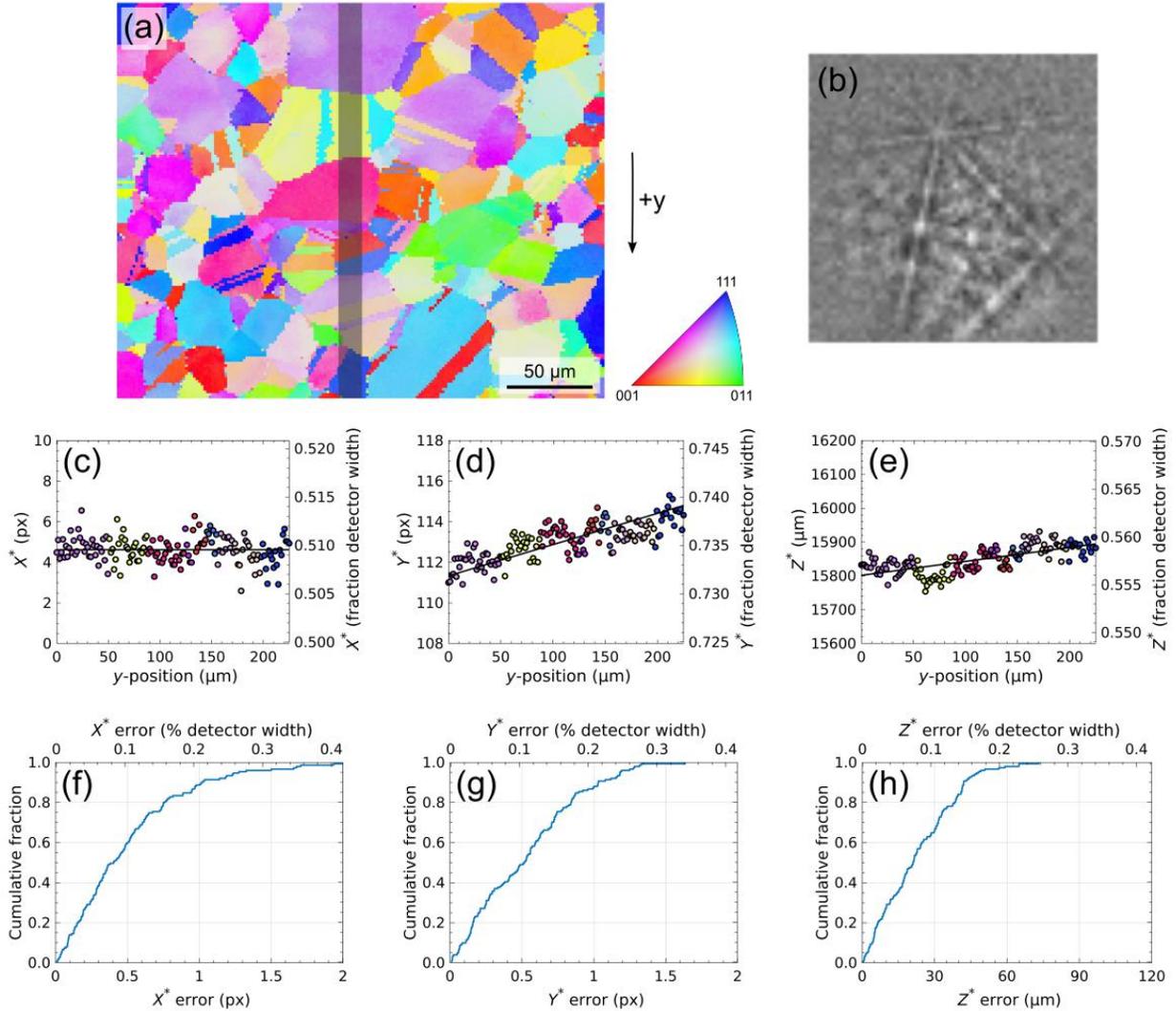

**Fig. 7.** Performance of the SNOBFIT algorithm in determining pattern centers on an experimental Ni dataset. (a) Inverse pole figure (IPF) map highlighting the region over which the linescans in parts c-e were taken. (b) Typical 8×8 binned pattern (60×60 px) from the dataset. (c-e) Pattern centers fitted along the *y*-direction, where each point represents the average pattern center over 10 patterns in the horizontal direction. Points are color coded according to the IPF map in part a. Lines represent the expected variation across the map due to the diffracting geometry. (f-h) Cumulative distribution of the absolute errors in parts c-e, assuming the line to be the true pattern center.

Although we cannot evaluate accuracy in this case (since we do not know the true pattern center), the correct detection of relative changes in pattern center across the sample gives confidence in the accuracy of the method. If we assume that the solid lines in Fig. 7c-e represent the true pattern center, the accuracy of the fitting procedure can be evaluated by considering the spread of the data points about this line. Fig. 7f-h display the cumulative distribution of absolute errors of these data points from the solid line. Because we do not know the true pattern center, these values represent a lower bound for the error. Nonetheless, the results show that for $X^*$ and $Y^*$, 90% of points have an error less than 0.2% detector width whereas for $Z^*$, 90% of points have an error less than 0.15%. This shows that $Z^*$ is the pattern center component that is most accurately fitted, in agreement with our results from Fig. 6 and previous studies [17,19,36]. This is expected since $Z^*$ cannot be easily compensated by a crystal rotation like $X^*$ and $Y^*$. These errors are about five times larger than those obtained for simulated patterns in Fig. 6a-c, which show errors of 0.05% for $X^*$ and $Y^*$ and 0.03% for $Z^*$ for 8×8 binned patterns with moderate noise.

One possible reason for the additional error in the experimental data is an orientation dependence of the pattern center error. In Fig. 7c-e, points are color coded according to the IPF map in Fig. 7a to show which grain each data point originates from, and the data show that the pattern center errors are non-randomly distributed about the drawn line. In fact, the fitted pattern centers appear to show a marked grain dependence. For example, the yellow data points near $y = 70$ μm fall mostly above the line for $Y^*$ but below the line for $Z^*$, and the opposite trend appears for the cream colored points near $y = 180$ μm. Similar patterns can also be seen for other grains. We note that in Fig. 6, the average pattern center was computed from 10 distinct orientations whereas in Fig. 7, the 10 patterns were generally from the same grain, except near grain boundaries. In fact, the data in Fig. 7 shows a lower spread near the grain boundaries where patterns from multiple grains are used, which confirms an orientation dependent systematic error in the fitted pattern center. These findings agree with our previous observations about the optimization landscape, where the global maximum is shifted from the true pattern center depending on the orientation in question. Thus, for the best possible accuracy, the pattern center should not only be averaged over numerous patterns as previously recommended by other authors [14,15,18,32,36], but these patterns should ideally be taken from different grains to reduce systematic errors due to orientation dependence. These systematic errors arise from orientation dependent slop, since the locations of the bands within a pattern affect how easily pattern shifts can be accommodated by crystal rotations. This may present difficulties for certain datasets, such as those involving single crystal specimens, which have often been used to study the accuracy of pattern center determination algorithms [14,32,34].

There are likely also additional error sources in the experimental data. The simulated patterns were strain-free, whereas the experimental patterns undoubtedly contained some finite amount of strain, especially near grain boundaries. In addition, a significant number of patterns used were near or even on grain boundaries, which produces lower quality EBSD patterns and sometimes even overlapping patterns. We have also not taken into account optical distortions by the detector. It also has previously been reported that hardware binning is considerably worse than software binning in terms of degrading information quality [52], and thus our results in Fig. 6 could be underestimating the error introduced by binning. These error sources are all likely present to some degree and contribute to the reduced pattern center accuracy in the experimental data. Future work should investigate ways to reduce these sources of error to further improve the accuracy of pattern center determination.

## 6. Conclusions

In this work, we have considered in detail the problem of pattern center determination in EBSD by comparison to dynamically simulated patterns. More specifically, we have focused on the approach where the pattern center and orientation are simultaneously fitted by direct optimization in six-dimensional space. Our results reveal that there is significant slop between pattern center and orientation and that surprisingly large pattern shifts can be almost completely accommodated by an appropriate rotation. In addition, the optimization landscape is noisy, which makes this a challenging landscape to optimize using standard search algorithms.

To effectively locate the global optimum in this noisy and sloppy landscape, we have applied the SNOBFIT algorithm, a global search algorithm designed for noisy objective functions, and have found that it successfully finds the global optimum. We have compared these results to the popular Nelder-Mead simplex algorithm, which we demonstrate easily gets trapped in local optima, thus giving considerably less accurate pattern center estimates. However, we find that even the pattern center accuracy from the SNOBFIT algorithm is limited by the nature of the optimization landscape, where the combined effects of slop and noise can displace the global optimum from the true pattern center. This imposes a practical limit on the accuracy of pattern center determination for a single pattern, which we have found to be ~0.05% detector width for pristine 480×480 px simulated patterns.

However, by averaging the pattern center from 10 distinct orientations, the error can be reduced to less than 0.008% for simulated patterns; the combined use of multiple pattern fitting and global optimization appears to be very promising for highly accurate EBSD. We have also studied the effect of binning and noise on the potential accuracy of this method. As expected, the error generally increases with noise and binning, but this method appears to be surprisingly tolerant to binning and noise.

We then applied the SNOBFIT algorithm to an experimental Ni dataset with aggressively binned 60×60 px patterns and successfully determined relative changes in pattern center across the map, demonstrating the effectiveness of the present fitting routine even for noisy, highly-binned experimental patterns. These results also suggest an orientation dependence, where certain grains lead to a systematic error in pattern center due to the slop and noise present in the optimization landscape. We therefore recommend that pattern center estimates should be averaged across numerous grains for the best possible accuracy. Source code for our pattern center fitting implementations are available at: https://github.com/epang22/pcglobal.

**Acknowledgements**

The authors would like to thank Prof. Marc De Graef (Carnegie Mellon University) for assistance with the dictionary indexing method and for providing scripts to interface with EMsoft. E.L.P. is supported by the NSF Graduate Research Fellowship Program under grant number DGE-1745302. P.M.L. is supported by grant number 7026-00126B from the Danish Council for Independent Research.

**References**

[1] M.C. Demirel, B.S. El-Dasher, B.L. Adams, A.D. Rollett, Studies on the accuracy of electron backscatter diffraction measurements, in: A.J. Schwartz, M. Kumar, B.L. Adams (Eds.), Electron Backscatter Diffr. Mater. Sci., 1st ed., Springer, New York, 2000: pp. 65–74.
[2] I. Brough, P.S. Bate, F.J. Humphreys, Optimising the angular resolution of EBSD, Mater. Sci. Technol. 22 (2006) 1279–1286.
[3] S. Wright, M. Nowell, High-speed EBSD, Adv. Mater. Process. 166 (2008) 29–32.
[4] F.J. Humphreys, Quantitative metallography by electron backscattered diffraction, J. Microsc. 195 (1999) 170–185.

[5]  F.J. Humphreys, Review: Grain and subgrain characterisation by electron backscatter diffraction, J. Mater. Sci. 36 (2001) 3833–3854.

[6]  G. Nolze, R. Hielscher, A. Winkelmann, Electron backscatter diffraction beyond the mainstream, Cryst. Res. Technol. 52 (2017) 1600252.

[7]  Y.H. Chen, S.U. Park, D. Wei, G. Newstadt, M.A. Jackson, J.P. Simmons, M. De Graef, A.O. Hero, A Dictionary Approach to Electron Backscatter Diffraction Indexing, Microsc. Microanal. 21 (2015) 739–752.

[8]  M.A. Jackson, E. Pascal, M. De Graef, Dictionary Indexing of Electron Back-Scatter Diffraction Patterns: a Hands-On Tutorial, Integr. Mater. Manuf. Innov. 8 (2019) 226–246.

[9]  A. Winkelmann, C. Trager-Cowan, F. Sweeney, A.P. Day, P. Parbrook, Many-beam dynamical simulation of electron backscatter diffraction patterns, Ultramicroscopy. 107 (2007) 414–421.

[10] P.G. Callahan, M. De Graef, Dynamical Electron Backscatter Diffraction Patterns. Part I: Pattern Simulations, Microsc. Microanal. 19 (2013) 1255–1265.

[11] S. Singh, F. Ram, M. De Graef, Application of forward models to crystal orientation refinement, J. Appl. Crystallogr. 50 (2017) 1664–1676.

[12] G. Nolze, M. Jürgens, J. Olbricht, A. Winkelmann, Improving the precision of orientation measurements from technical materials via EBSD pattern matching, Acta Mater. 159 (2018) 408–415.

[13] S.I. Wright, M.M. Nowell, S.P. Lindeman, P.P. Camus, M. De Graef, M.A. Jackson, Introduction and comparison of new EBSD post-processing methodologies, Ultramicroscopy. 159 (2015) 81–94.

[14] D. Fullwood, M. Vaudin, C. Daniels, T. Ruggles, S.I. Wright, Validation of kinematically simulated pattern HR-EBSD for measuring absolute strains and lattice tetragonality, Mater. Charact. 107 (2015) 270–277.

[15] B.E. Jackson, J.J. Christensen, S. Singh, M. De Graef, D.T. Fullwood, E.R. Homer, R.H. Wagoner, Performance of Dynamically Simulated Reference Patterns for Cross-Correlation Electron Backscatter Diffraction, Microsc. Microanal. 22 (2016) 789–802.

[16] F. Ram, S. Wright, S. Singh, M. De Graef, Error analysis of the crystal orientations obtained by the dictionary approach to EBSD indexing, Ultramicroscopy. 181 (2017) 17–26.

[17] N.C. Krieger-Lassen, J.B. Bilde-Sorensen, Calibration of an electron back-scattering pattern set-up, J. Microsc. 170 (1993) 125–129.

[18] N.C. Krieger Lasssen, Automated Determination of Crystal Orientations from Electron Backscattering Patterns, Technical University of Denmark, 1994.

[19] T.B. Britton, C. Maurice, R. Fortunier, J.H. Driver, A.P. Day, G. Meaden, D.J. Dingley, K. Mingard, A.J. Wilkinson, Factors affecting the accuracy of high resolution electron backscatter diffraction when using simulated patterns, Ultramicroscopy. 110 (2010) 1443–1453.

[20] A.J. Wilkinson, D.J. Dingley, G. Meaden, Strain Mapping Using Electron Backscatter Diffraction, in: A.J. Schwartz, M. Kumar, B.L. Adams, D.P. Field (Eds.), Electron Backscatter Diffr. Mater. Sci., 2nd ed., Springer, New York, 2009: pp. 231–249.

[21] S. I. Wright, M. M. Nowell, D. P. Field, A Review of Strain Analysis Using Electron Backscatter Diffraction, Microsc. Microanal. 17 (2011) 316–329.

[22] M. De Graef, W.C. Lenthe, N. Schäfer, T. Rissom, D. Abou-Ras, Unambiguous Determination of Local Orientations of Polycrystalline CuInSe 2 Thin Films via Dictionary-Based Indexing, Phys. Status Solidi - Rapid Res. Lett. 1900032 (2019).

[23] E.L. Pang, C.A. McCandler, C.A. Schuh, Reduced cracking in polycrystalline ZrO2-CeO2 shape-memory ceramics by meeting the cofactor conditions, Acta Mater. 177 (2019) 230–239.

[24] J. Hjelen, R. Ørsund, E. Hoel, P. Runde, T. Furu, E. Nes, EBSP, Progress in Technique and

[25] D.A. Carpenter, J.L. Pugh, G.D. Richardson, L.R. Mooney, Determination of pattern centre in EBSD using the moving-screen technique, J. Microsc. 227 (2007) 246–247.

[26] C. Maurice, K. Dzieciol, R. Fortunier, A method for accurate localisation of EBSD pattern centres, Ultramicroscopy. 111 (2011) 140–148.

[27] S. Biggin, D.J. Dingley, A general method for locating the X-ray source point in Kossel diffraction, J. Appl. Crystallogr. 10 (1977) 376–385.

[28] J.A. Venables, R. Bin-Jaya, Accurate microcrystallography using electron back-scattering patterns, Philos. Mag. 35 (1977) 1317–1332.

[29] K. Mingard, A. Day, C. Maurice, P. Quested, Towards high accuracy calibration of electron backscatter diffraction systems, Ultramicroscopy. 111 (2011) 320–329.

[30] D.J. Dingley, K. Baba-Kishi, Use of electron back scatter diffraction patterns for determination of crystal symmetry elements, Scan. Electron Microsc. 2 (1986) 383–391.

[31] S.I. Wright, Individual lattice orientation measurements: Development and application of a fully automated technique, Yale University, 1992.

[32] J. Alkorta, M. Marteleur, P.J. Jacques, Improved simulation based HR-EBSD procedure using image gradient based DIC techniques, Ultramicroscopy. 182 (2017) 17–27.

[33] T. Tanaka, A.J. Wilkinson, Pattern matching analysis of electron backscatter diffraction patterns for pattern centre, crystal orientation and absolute elastic strain determination – accuracy and precision assessment, Ultramicroscopy. 202 (2019) 87–99.

[34] J. Basinger, D. Fullwood, J. Kacher, B. Adams, Pattern Center Determination in Electron Backscatter Diffraction Microscopy, Microsc. Microanal. 17 (2011) 330–340.

[35] J. Basinger, Detail Extraction from Electron Backscatter Diffraction Patterns, Brigham Young University, 2011.

[36] A. Day, Developments in the EBSP Technique and their Application to Grain Imaging, University of Bristol, 1993.

[37] A. Winkelmann, Dynamical Simulation of Electron Backscatter Diffraction Patterns, in: A.J. Schwartz, M. Kumar, B.L. Adams, D.P. Field (Eds.), Electron Backscatter Diffr. Mater. Sci., 2nd ed., Springer, 2009: pp. 21–33.

[38] F. Ram, S. Zaefferer, T. Jäpel, D. Raabe, Error analysis of the crystal orientations and disorientations obtained by the classical electron backscatter diffraction technique, J. Appl. Crystallogr. 48 (2015) 797–813.

[39] O. Engler, V. Randle, Introduction to texture analysis: macrotexture, microtexture and orientation mapping, 2nd ed., CRC Press, 2010.

[40] E.A. Owen, E.L. Yates, XLI. Precision measurements of crystal parameters, London, Edinburgh, Dublin Philos. Mag. J. Sci. 15 (1933) 472–488.

[41] L.M. Peng, G. Ren, S.L. Dudarev, M.J. Whelan, Debye-Waller factors and absorptive scattering factors of elemental crystals, Acta Crystallogr. Sect. A Found. Crystallogr. 52 (1996) 456–470.

[42] R.N. Gutenkunst, J.J. Waterfall, F.P. Casey, K.S. Brown, C.R. Myers, J.P. Sethna, Universally sloppy parameter sensitivities in systems biology models, PLoS Comput. Biol. 3 (2007) 1871–1878.

[43] B.K. Mannakee, A.P. Ragsdale, M.K. Transtrum, R.N. Gutenkunst, Sloppiness and the Geometry of Parameter Space, in: L. Geris, D. Gomez-Cabrero (Eds.), Uncertain. Biol., Springer, 2016: pp. 271–299.

[44] J.C. Lagarias, J.A. Reeds, M.H. Wright, P.E. Wright, Convergence Properties of the Nelder-Mead Simplex Method in Low Dimensions, SIAM J. Optim. 9 (1998) 112–147.

[45] K.I.M. McKinnon, Convergence of the Nelder-Mead Simplex Method to a Nonstationary Point, SIAM J. Optim. 9 (1998) 148–158.

Applications, Textures Microstruct. 20 (1993) 29–40.

[46] L.M. Rios, N. V. Sahinidis, Derivative-free optimization: A review of algorithms and comparison of software implementations, J. Glob. Optim. 56 (2013) 1247–1293.

[47] L. Han, M. Neumann, Effect of dimensionality on the Nelder-Mead simplex method, Optim. Methods Softw. 21 (2006) 1–16.

[48] W. Huyer, A. Neumaier, SNOBFIT - Stable Noisy Optimization by Branch and Fit, ACM Trans. Math. Softw. 35 (2008) 1–25.

[49] A. Neumaier, SNOBFIT, (n.d.).

[50] T. Friedrich, A. Bochmann, J. Dinger, S. Teichert, Application of the pattern matching approach for EBSD calibration and orientation mapping, utilising dynamical EBSP simulations, Ultramicroscopy. 184 (2018) 44–51.

[51] J. Alkorta, Limits of simulation based high resolution EBSD, Ultramicroscopy. 131 (2013) 33–38.

[52] T.B. Britton, J. Jiang, R. Clough, E. Tarleton, A.I. Kirkland, A.J. Wilkinson, Assessing the precision of strain measurements using electron backscatter diffraction - part 1: Detector assessment, Ultramicroscopy. 135 (2013) 126–135.